\documentclass[12pt]{iopart}
\usepackage{graphicx}
\begin{document}

\title{Hybrid simulations of mini-magnetospheres in the laboratory}

\author{L. Gargat\'{e}}
\address{GoLP/IPFN, Instituto Superior T\'{e}cnico, Av. Rovisco Pais, 1049-001 Lisbon, Portugal}
\ead{luisgargate@ist.utl.pt}

\author{R. Bingham}
\address{Space Science \& Technology Dept,Science \& Technology Facilities Council, Rutherford Appleton Laboratory, Harwell Science and Innovation Campus, Didcot, Oxon, OX11 0QX UK}

\author{R. A. Fonseca}
\address{DCTI, Instituto Superior de Cincias do Trabalho e da Empresa, Av. For\,cas Armadas, 1649-026 Lisbon, Portugal}
\address{GoLP/IPFN, Instituto Superior T\'{e}cnico, Av. Rovisco Pais, 1049-001 Lisbon, Portugal}

\author{R. Bamford}
\address{Space Science \& Technology Dept,Science \& Technology Facilities Council, Rutherford Appleton Laboratory, Harwell Science and Innovation Campus, Didcot, Oxon, OX11 0QX UK}

\author{A. Thornton}
\address{University of Manchester,  Sackville Street Building, Manchester, M60 1QD, U.K}

\author{K. Gibson}
\address{University of Manchester,  Sackville Street Building, Manchester, M60 1QD, U.K}

\author{J. Bradford}
\address{Space Science \& Technology Dept,Science \& Technology Facilities Council, Rutherford Appleton Laboratory, Harwell Science and Innovation Campus, Didcot, Oxon, OX11 0QX UK}

\author{L. O. Silva}
\address{GoLP/IPFN, Instituto Superior T\'{e}cnico, Av. Rovisco Pais, 1049-001 Lisbon, Portugal}

\begin{abstract}
Solar energetic ions are a known hazard to both spacecraft electronics and to manned space flights in interplanetary space missions that extend over a long period of time. A dipole-like magnetic field and a plasma source, forming a mini magnetosphere, are being tested in the laboratory as means of protection against such hazards. We investigate, via particle-in-cell hybrid simulations, using kinetic ions and fluid electrons, the characteristics of the mini magnetospheres. Our results, for parameters identical to the experimental conditions, reveal the formation of a mini-magnetosphere, whose features are scanned with respect to the plasma density, the plasma flow velocity, and the intensity of the dipole field. Comparisons with a simplified theoretical model reveal a good qualitative agreement and excellent quantitative agreement for higher plasma dynamic pressures and lower B-fields. 
\end{abstract}

%Uncomment for PACS numbers title message
\pacs{94.05.-a,94.05.Hk,94.05.sj,94.30.Bg,52.65.Ww,52.65.Rr}
% Keywords required only for MST, PB, PMB, PM, JOA, JOB? 
%\vspace{2pc}
%\noindent{\it Keywords}: Article preparation, IOP journals
% Uncomment for Submitted to journal title message
%\submitto{\JPA}
% Comment out if separate title page not required
\maketitle

\section{Introduction}

The space environment is hazardous both to spacecrafts (e.g. electronics, external panels) as well as to the astronauts, not only due to radiation (e.g. UV radiation, gamma rays) but also due to very energetic charged particles originating from the Sun. These charged particles are mainly protons, generally referred to as Solar Energetic Particles (SEP), and can reach energies up to hundreds of MeV. 
Much in the same way that the Earth's magnetosphere deflects energetic charged particles, thus acting as a shield, it is also expected that by creating and sustaining a mini-magnetosphere around a spacecraft it is possible to protect the spacecraft. Emulating the Earth's magnetosphere, it can be envisioned to produce mini-magnetosphere using a dipole magnetic field and a plasma source. 

Similar configurations have been recently studied in the context of spacecraft propulsion in the Solar System \cite{winglee,mendonca,loureiro}. One of the critical issues identified in these studies is the expansion of the plasma, injected near the dipole field; the propulsion scheme is only efficient whenever a magnetoshpere is created, as the expanding plasma drags the magnetic field and the field intensity decays slower than in the dipole field configuration ($\propto r^{-3}$). Detailed numerical simulations of the expansion of a magnetic bubble have also been performed recently \cite{winske, haibintang}, in the absence of a flowing plasma. The overall structure size, in the dimension perpendicular to the plasma flow, is in the order of $15\,\mathrm{km}$ to $20\,\mathrm{km}$ for efficient propulsion. 

SEP have energies in the MeV range, and a full study of how these particles are deflected by a magnetosphere has not yet been completed. Even in the well studied case of the Earth, a full understanding of how effectively SEPs are deflected is lacking; it is known that some very energetic particles (tens of MeVs) penetrate the magnetopause and are trapped in the EarthÕs radiation belt for very long periods. Physically, we can state that for deflection to be efficient the typical ion Larmor radius must be smaller than the typical size of the magnetic field structure.

The demonstration of the controlled generation of mini-magnetospheres has motivated laboratory experiments currently underway \cite{ziemba,winglee2,funaki,bamford}. In these experiments, a plasma beam, guided by an axial magnetic field in a cylindrical linear chamber, hits a dipole magnetic field created by a permanent magnet. Preliminary experimental results already reveal the formation of a very sharp shock structure, thus indicating the formation of a mini-magnetosphere (cf. Fig. 1).

In order to provide further insight into the underlying physics, and to guide future experimental developments, we have preformed simulations of the interaction of a plasma flow with a static dipole-like field, using  particle-in-cell \cite{dawson,birdsall} hybrid simulations using the code \textit{dHybrid} \cite{gargate}, with the visualization performed with the osiris framework \cite{fonseca}. In order to understand the key parameters that determine the mini-magnetosphere features, a parameter scan of the plasma density, the flow velocity and the dipole magnetic field intensity has been performed, giving a particular emphasis to the shock behavior (e.g. the distance of magnetopause to the dipole field origin). The simulations are compared with a simplified theoretical model, thus providing the required framework not only to understand and to optimize present day experiments but also to link the simulation and experimental work to the space plasma environment. It was found that the magnetopause distance to the dipole origin increases with increasing magnetic field intensity of the dipole and with decreasing kinetic pressure of the flowing plasma, as expected from theory.

The paper is outlined as follows. In Section II, we present the hybrid model used in the simulations, and the simplified theoretical framework that provides a physical picture for the most relevant feature of the mini-magnetosphere. In Section III, the simulation results are compared with the theoretical predictions, and a detailed discussion of the behavior of the system with the plasma parameters and the magnetic field is presented. Finally, we state the conclusions.

\section{Theoretical Framework}
\indent

Since the typical time and the length scales in the mini-magnetospheres in the laboratory are large, three-dimensional kinetic particle-in-cell simulations, where both ions and electrons are treated kinetically are not feasible. We resort instead to hybrid simulations, using the code \textit{dHybrid} \cite{gargate}. Hybrid models are commonly used in many problems in plasma physics (for a review see, for instance, ref. \cite{lipatov}). The hybrid set of equations is derived neglecting the displacement current in Amp\`{e}re's Law, considering quasi-neutrality and calculating moments of the Vlasov equation for the electrons in order to obtain the generalized Ohm's Law. In \textit{dHybrid}, the electron mass, the resistivity and the electron pressure are not considered; thus, the electric field is simply given by  $\vec{E}=-\vec{V}_e\times\vec{B}$, which can also be expressed as 
\begin{equation}
\vec{E}=-\vec{V}_i \times\vec{B}+\frac{1}{n}\left(\nabla\times\vec{B}\right)\times\vec{B}
\label{eq:efield}
\end{equation}
where we have written $\vec{V}_e= -\vec{J}/(|e| n) +\vec{V}_i = -(\nabla \times \vec{B})/(|e| n)+\vec{V}_i$, where $\vec{V}_i=\frac{1}{n}\int f_i\,\vec{v}\, d\vec{v}$ is the ion fluid velocity, and $n$ is the electron/ion density. Normalized simulation units are used: time is normalized to  $1/\omega_{ci}$, space is normalized to $c/\omega_{pi}$, charge is normalized to the proton charge $|e|$, and mass is normalized to the proton mass, where $\omega_{ci}$ is the ion cyclotron frequency and $\omega_{pi}$ is the ion plasma frequency. The magnetic field is advanced in time through Faraday's Law $\frac{\partial \vec{B}}{\partial t}=-\nabla\times\vec{E}$, with $\vec{E}$ calculated from Eq. (\ref{eq:efield}). In \textit{dHybrid}, the ions are treated as kinetic particles, and their velocity is updated via the Lorentz force. 

The hybrid approximation is clearly valid here, since the formation of the shock occurs on much longer time scales than the electron time scale. For the parameters considered, the electron Larmor radius is at most $1\%$ of the typical scale length considered in the simulations. Furthermore, it is the deflection of ions away from the spacecraft that is of key interest since ions are more difficult to deflect. Previous simulations for similar scenarios \cite{chapman,bingham,harold,kazeminezhad,delamere,nairn} have shown that  finite ion Larmor radius effects are important to capture the dynamics of the magnetosphere, specially in the experimental scenarios we are considering where the ion Larmor radius can be comparable to the size of the mini-magnetosphere; only a fully kinetic treatment of the ions can properly take into account these effects. 

A simple theoretical model, first considered for the Earth's magnetosphere, can provide insight into the location of the magnetopause with respect to the source of the dipole field \cite{bspp}. To derive the distance, $r_\mathrm{mp}$, from the nose of the magnetopause to the dipole field origin, it is considered that the total pressure is conserved across the magnetopause, that is, $\left[p+B^2/2\right]=0$, where $p$ is the plasma pressure, the second term in the left hand side is the magnetic field pressure, and the square brackets define differences between quantities upstream and downstream of the magnetopause. Upstream of the magnetopause, the plasma dynamic pressure dominates, and downstream of the magnetopause, the magnetic dipole field pressure dominates. In this case, $r_\mathrm{mp}$ is given by
\begin{equation}
r_\mathrm{mp}=\left(\frac{K\,B^2}{2\,n\,m_i\,v^2}\right)^{1/6}
\label{eq:scalinglaw}
\end{equation}
where $B$ is the magnetic field intensity at the edge of the magnet, where $n$ is the density, $m_i$ is the ion mass (protons), and $v$ is the flow velocity of the plasma. The parameter $K$ is a free parameter of the theory, accounting for both deviations of the magnetic field from its dipolar value at $r_\mathrm{mp}$, due to plasma currents, and $\kappa$, an efficiency coefficient in the plasma dynamic pressure equation $p_{dyn}=\kappa\,n\,\,m_i\,v^{2}$, that accounts for the non-ideal specular reflection of the particles when hitting the magnetopause. The value for $K$ is adjusted from simulation data and, neglecting the deviations from the magnetic field from its dipolar value at $r_\mathrm{mp}$ allows us to estimate $\kappa$, with $\kappa=1$ meaning specular reflection of particles off the magnetosphere. 

The magnetopause distance from the dipole origin, given by Eq. (\ref{eq:scalinglaw}), then depends on the plasma density $n$, the plasma velocity $v$ and the dipole magnetic field intensity. It is now possible to compare the simulation results from \textit{dHybrid} with this simple theoretical model, namely by comparing the plasma deflection distance at the magnetopause in the $x$ direction, as measured in the simulations,  with the predictions of Eq. (\ref{eq:scalinglaw}).

In order to understand the similarities and the differences between the well studied case of the Earth's magnetosphere and the laboratory mini magnetospheres considered here, it is instructive to examine the typical parameters in both scenarios. In the space plasma environment near the Earth, the solar wind, primarily protons and electrons, has a typical ion density of $n\sim5\,\mathrm{cm^{-3}}$, a typical ion velocity of $v\sim450\,\mathrm{km/s}$, a typical ion temperature of $T\sim20\,\mathrm{eV}$, and the Interplanetary Magnetic Field (IMF) is $B\sim10\,\mathrm{nT}$. These parameters yield an acoustic Mach number $M_{cs}\sim7.3$, an Alfv\`enic Mach number $M_{ca}\sim4.6$, a plasma $\beta\sim0.4$, an ion inertial length $c/\omega_{pi}\sim102\,\mathrm{km}$, and an ion Larmor radius $r_{L}\sim469\,\mathrm{km}$ for protons. The solar wind drags the frozen-in IMF and forms a super Alfv\`enic shock when hitting the magnetopause, creating the Earth's bow shock. The ion Larmor radius is very small compared to the size of the Earth's magnetosphere; a multi-fluid or MHD approach are reasonable approximations to tackle this kind of interaction. 

For the simulations of the laboratory experiments, the proton/electron plasma flow has a typical ion density of $n\sim10^{12}\,\mathrm{cm^{-3}}$, a typical ion velocity of $v\sim400\,\mathrm{km/s}$, a typical ion temperature of $T\sim5\,\mathrm{eV}$, and the guiding magnetic field (in a parallel IMF configuration) is $B\sim0.02\,\mathrm{T}$. These parameters correspond to an acoustic Mach number $M_{cs}\sim12.9$, an Alfv\`enic Mach number $M_{ca}\sim0.9$, $\beta\sim0.005$, an ion inertial length $c/\omega_{pi}\sim22.8\,\mathrm{cm}$, and an ion Larmor radius $r_{L}\sim20.8\,\mathrm{cm}$ for protons. Here, the ion Larmor radius is of the order of the magnetosphere being formed.

Unlike the Earth scenario, here the shock is not super Alfv\`enic and the $\beta$ is much lower: i) the shock formed by the plasma flow against the magnetosphere has slightly different characteristics from the Earth's bow shock e.g. a magnetosheath is still observed, but the plasma density does not increase from the solar wind to the magnetosheath), and ii) the plasma flow follows the magnetic field lines, and the magnetic field topology near the magnet is not significantly modified, thus keeping the dipole like structure.  

\section{Simulation Results}
\indent

In order to make connection with the experiments now in progress, we will use the laboratory parameters for the mini-magnetospheres experiments described in the previous section. In the experimental setup, a hydrogen plasma is created inside a cylindrical linear chamber and confined with an axial magnetic field; a cylindrically shaped magnet is positioned in the plasma path. The magnetic moment of the dipole like field created by the magnet is perpendicular to the plasma propagation direction \cite{bamford}.
\begin{figure}
\begin{center}
\includegraphics[width=8cm]{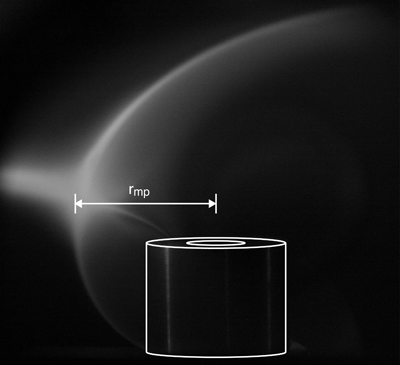}
\caption{The mini-magnetosphere in the laboratory. The permanent magnet is digitally highlighted in the image. The plasma is coming from the left-hand side.}
\end{center}
\label{fig:experiment}
\end{figure}

In the simulations, the plasma is moving in the $+x$ direction and occupies all the simulation box. An external dipole magnetic field, whose center is located in the middle of the simulation box, is imposed with the magnetic moment aligned along the $+z$ direction, mimicking the permanent magnet, along with a lower intensity constant magnetic field on all the simulation box in the $+x$ direction to account for the effects of the axial confining magnetic field used in the experiment. 
\begin{figure}
\begin{center}
\includegraphics[width=8cm]{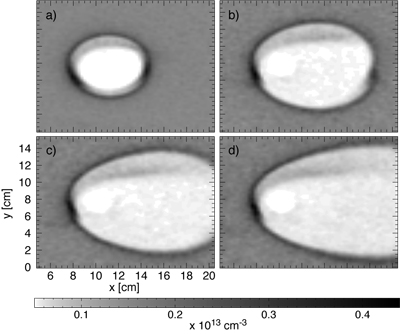}
\caption{Cut along the $xy$ plane for $z=7.64\,\mathrm{cm}$ (middle of the simulation box) of the plasma density at times a) $24.6\,\mathrm{ns}=0.047\,\mathrm{\omega_{ci}^{-1}}$, b) $73.9\,\mathrm{ns}=0.142\,\mathrm{\omega_{ci}^{-1}}$, c) $123.2\,\mathrm{ns}=0.236\,\mathrm{\omega_{ci}^{-1}}$, d) $172.5\,\mathrm{ns}=0.331\,\mathrm{\omega_{ci}^{-1}}$, for a plasma density of $10^{18}\,\mathrm{m^{-3}}$, a plasma temperature of $T=5\,\mathrm{eV}$, a plasma flow velocity of $620\,\mathrm{km/s}$, and a magnetic field intensity of $0.2\,\mathrm{T}$, corresponding to $M_{cs}=20$, $M_{ca}=1.4$ and $\beta=0.005$.}
\end{center}
\label{fig:densevo}
\end{figure}

The simulations discussed here are performed in a 3D simulation box, with a box size of $20.35\,\mathrm{cm}$ in the $x$ direction, $15.27\, \mathrm{cm}$ in the $y$ and $z$ directions and $80\times 60\times 60$ cells in the $x$, $y$ and $z$ directions, respectively, corresponding to a box size of $0.98\times0.73\times0.73\,r_\mathrm{L}^3=0.89\times0.67\times0.67\,\left(\mathrm{c/\omega_{pi}}\right)^3$. Simulations are run up to $t=0.38\,\mathrm{\omega_{ci}^{-1}}\sim211\,\mathrm{\omega_{pi}^{-1}}$ for most cases, and up to $t=1.13\,\mathrm{\omega_{ci}^{-1}}\sim632\,\mathrm{\omega_{pi}^{-1}}$ to check the stability of the mini-magnetosphere. Normalizing quantities are $n_0=1\times10^{18}\,\mathrm{m^{-3}}$, $M_{ca}=436.58\,\mathrm{km/s}$ and $c/\omega_{pi}=22.8\,\mathrm{cm}$.
\begin{figure}
\begin{center}
\includegraphics[width=10cm]{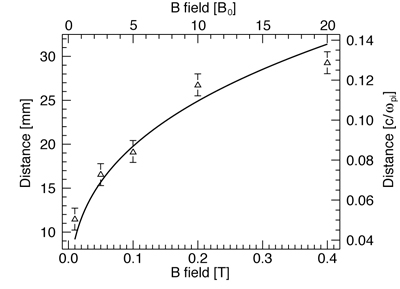}
\caption{Distance of the mangetopause to the dipole origin, $r_\mathrm{mp}$, as a function of the magnetic field intensity at the edge of the permanent magnet. The solid line represents the theoretical prediction, Eq. (\ref{eq:scalinglaw}). The triangles represent the values measured in the simulations, with the error bars describing the resolution of the simulation.}
\end{center}
\label{fig:radiusvsb}
\end{figure}

Since one of the key observables in the experiments is the distance, directly facing the plasma flow, from the magnet to the magnetopause, $r_\mathrm{mp}$, we have performed a parameter scan in order to understand how $r_\mathrm{mp}$ depends on the magnetic field intensity of the dipole, the plasma density of the incoming flow, and the velocity of the plasma flow.
\begin{figure}
\begin{center}
\includegraphics[width=10cm]{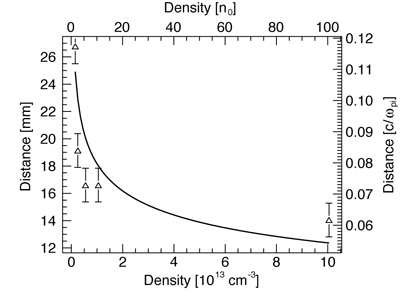}
\caption{Dependence of the distance of the magnetopause to the dipole origin, $r_\mathrm{mp}$, on the density of the plasma flow, for the baseline simulation parameters. The solid line represents the theoretical prediction, Eq. (\ref{eq:scalinglaw}). The triangles denote the simulation results, with the error bars describing the resolution of the simulation.}
\end{center}
\label{fig:radiusvsdens}
\end{figure}

In all the scenarios, as soon as the simulation starts, the plasma is expelled from the most intense magnetic field region, near the dipole origin (cf.  Fig. 2). It should be noticed that the distance from the magnetopause to the dipole origin, measured along the $x$ direction and passing through the dipole origin, does not change significantly in time, reaching the equilibrium value on the ion response time scale, as illustrated in the Fig. 2, where the formation of the shock can also be followed. For all the simulation scenarios the value for $K$, in Eq. (\ref{eq:scalinglaw}), was $K=6.09\times10^{-12}\,\mathrm{m^6}$. Neglecting deviations of the magnetic field from its assumed dipolar value, a value for $\kappa\sim1$ can be estimated, meaning particles are specularly reflected. We have checked teh assumption that the magnetic field does not deviate significantly from a dipole configuration; in our simulations the deviations are below $7.3\%$. For the Earth's case, assuming a magnetic moment of $8.05\times10^{22}\,\mathrm{A\,m^2}$, $\kappa\sim0.29$. This means that the mini-magnetospheres represent ideal setting to test and to probe the theoretical models. 
\begin{figure}
\begin{center}
\includegraphics[width=10cm]{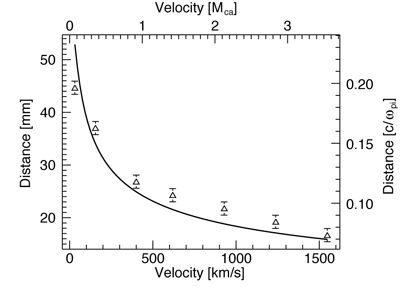}
\caption{Distance of the magnetopause to the dipole origin, $r_\mathrm{mp}$, as a function of the plasma flow velocity, for the baseline simulation parameters. The solid line represents the theoretical prediction, Eq. (\ref{eq:scalinglaw}). The triangles denote the simulation results, with the error bars describing the resolution of the simulation.}
\end{center}
\label{fig:radiusvsvel}
\end{figure}

In the first simulation scenario, we have scanned the magnetic field intensity, in the parameter range to be available in the near future experimental campaigns, with the magnetic field on the edge of the magnet varying from $0.01\,\mathrm{T}$ to $0.4\,\mathrm{T}$; the initial plasma parameters were kept fixed, identical to the experimental scenario. In Fig. 3, $r_\mathrm{mp}$ measured in the simulation is compared with the theoretical prediction from Eq. (\ref{eq:scalinglaw}). Similar qualitative behavior is obtained, but quantitatively some discrepancies are visible; quantitative agreement is better for lower magnetic pressures. It should be stressed, however, that for the same parameters used in the experimental setup (at the reference B field of $0.2\,\mathrm{T}$) the measurement of $r_\mathrm{mp}$ in the simulation yields $26.7\,\mathrm{mm}$, with a resolution of $2.5\, \mathrm{mm}$ (corresponding to the cell size) which is in very good agreement with the experimental value of $28.5\,\mathrm{mm}$ \cite{bamford2,gargate2}, and provides an additional validation of the simulation results for the range of parameters of the experiments. 

Scanning the distance $r_\mathrm{mp}$ as a function of the plasma density, Fig. 4,  the qualitative behavior is recovered, with a slight deviation from the theoretical model, as in the previous scenario. For the plasma densities scanned, the $\beta$'s range from $\beta=0.005$ to $\beta=0.5$. The dependance $r_\mathrm{mp}\propto n^{-1/6}$, Eq. (\ref{eq:scalinglaw}), means that $r_\mathrm{mp}$ will have small changes with varying density, while for variations of magnetic field, $r_\mathrm{mp}\propto B^{1/3}$, and for variations of velocity, $r_\mathrm{mp}\propto v^{-1/3}$, which means that more significant changes of $r_\mathrm{mp}$ are observed in these cases for our parameter scan, and for our numerical parameters.

Finally, in the third scenario, the velocity of the solar wind was varied from $30.97\,\mathrm{km/s}$ up to $1548\,\mathrm{km/s}$, corresponding to acoustic Mach numbers from $M_{cs}=1$ to $M_{cs}=50$ and Alfv\`enic Mach numbers from $M_{ca}=0.07$ to $M_{ca}=3.5$. The results, depicted in Fig. 5, follow the same behavior as in the previous scenarios, with a small deviation from the theoretical values for most of the measured points. The discrepancies of measured values from the theoretical values are discussed in the next section, and are associated with the simplifying assumptions of the theoretical model.

\section{Discussion and Conclusions}
\indent

We have successfully showed the ability to perform hybrid simulations for parameters relevant for on-going mini-magnetosphere experiments. We have found that the distance from the magnetopause to the dipole origin in the mini magnetosphere follows the qualitative behavior of a simplified theoretical model. 

The quantitative discrepancies found are due to the assumptions of the model, in particular the fact that the thermal pressure is neglected, accounting for deviations for lower $v$ values in Fig. 5. If we include in Eq. (\ref{eq:scalinglaw}) the plasma thermal pressure, the maximum deviation from the theoretical model is below $15\%$. The electron dynamic ram pressure could also be considered in Eq. (\ref{eq:scalinglaw}), but due to the low mass of the electrons compared to the mass of the ions, it has a negligible effect, modifying $r_\mathrm{mp}$ by a factor of $\sim0.01\%$.

Furthermore, the Rankine-Hugoniot jump conditions are implicitly used in Eq. (\ref{eq:scalinglaw}). These are derived using an ideal one fluid magnetohydrodynamics model which has limited applicability in the mini-magnetosphere scenario; kinetic effects associated with finite Larmor radius effects should be important in this scenario \cite{gargate,kazeminezhad}. 

The differences of the experimental setup to the space plasma parameters should also be considered as a guide for further numerical and experimental explorations. The main difference resides in the number density of the solar wind $\approx 5\,cm^{-3}$ for the relevant cases, which represents a much lower value than the densities used here, accounting for the differences outlined before in the plasma $\beta$ and the Mach numbers. Assuming that the magnetic field intensities of a system on-board of a spacecraft can be of the order of magnitude of the ones tested here, the system should stop the incoming solar wind at longer $r_\mathrm{mp}$, according to Eq. (\ref{eq:scalinglaw}), by a factor of $\left(n/n_{sw}\right)^{1/6}\sim76$ resulting in a standoff distance of a few meters. 

In order to push the concept of the mini-magnetosphere further, the injection of plasma in the region of the dipole field to expand the magnetospheric magnetic field has to be considered. Plasma injection can lead to the expansion of the dipole magnetic field, thus changing the decay law from $1/r^3$ to $1/r^\eta$ with $\eta<3$, and increasing $r_\mathrm{mp}$, {\it i.e.} increasing the protective distance of such a system for a space application. An estimate of operating parameters for this configuration is calculated based on the requirement of deflecting $1\,\mathrm{MeV}$ protons and considering a magnetic field intensity decay of $r^{-1}$. For efficient reflection, we require the Larmor radius to be a fraction, $f\sim20\%$, of the distance of the proton to the spacecraft, yielding a magnetic field intensity of $0.72\,\mathrm{T}$ generated by a current loop with $r=1\,\mathrm{m}$, corresponding to a magnetic moment $M\sim7.2\times10^6\,\mathrm{A\,m^2}$. 

These estimates provide only a crude approximation; our results indicate that the requirements for the magnetic field intensity can be relaxed in the self-consistent configuration. As observed in our simulations, even when $r_{L}$ is comparable to $r_\mathrm{mp}$, the incoming plasma is totally deflected at a distance of the order of $r_\mathrm{mp}$. This will be explored in a future publication. 

Future work on the subject will then focus on the behavior of a mini-magnetosphere in a space plasma environment, accounting for the much lower densities of the solar wind. The injection of plasma and the expansion of the dipolar magnetic field will also be tested, in the presence of the solar wind. The deflection of energetic particles by these configurations will be considered, resorting to test particles, thus providing information that will allow for a detail assessment of the role of mini-magnetospheres in the space environment.

\end{document}